\documentclass[12pt]{article}
\usepackage{jheppub}
\pdfoutput=1
\usepackage{amsmath,bbm,array,amsfonts,graphicx,wrapfig,lscape,float,mathtools,multirow,longtable,amsthm}
\usepackage[dvipsnames]{xcolor}
\usepackage{stackrel}
\usepackage[all]{xy}
\usepackage{caption}
\usepackage{subcaption}
\usepackage[bottom]{footmisc}
\usepackage{mathrsfs}
\usepackage{tikz}
\usepackage{bm}
\usepackage{color}
\usepackage{enumerate}
\usetikzlibrary{decorations.pathreplacing}
\usepackage{diagbox}
\numberwithin{equation}{section}
\numberwithin{figure}{section}
\numberwithin{table}{section}
\usepackage{pgfplots}
\pgfplotsset{compat=1.14}
\usepackage{makecell}
\usepackage{lscape}
\usepackage[utf8]{inputenc}
\usepackage{mathtools}
\usepackage{epigraph}

\usepackage[autostyle=true]{csquotes}
\newtheorem{definition}{Definition}[section]
\newtheorem{theorem}{Theorem}[section]

\newtheorem{conjecture}[theorem]{Conjecture}

\newtheorem{question}{Question}

\usepackage[toc,page]{appendix}
\usepackage{tocloft}

	\title{Some Open Questions in Quiver Gauge Theory}
	
	\author[a]{Jiakang Bao,}
	\author[b]{Amihay Hanany,}
	\author[c,a,d,e]{Yang-Hui He,}
	\author[a]{Edward Hirst}
	
	\affiliation[a]{
		Department of Mathematics, City, University of London, EC1V 0HB, UK}
	\affiliation[b]{
		Theoretical Physics Group, The Blackett Laboratory, Imperial College London, Prince Consort Road, London, SW7 2AZ, UK}
	\affiliation[c]{
		London Institute of Mathematical Sciences, Royal Institution of GB, W1S 4BS, UK}
	\affiliation[d]{
		Merton College, University of Oxford, OX1 4JD, UK}
	\affiliation[e]{
		School of Physics, NanKai University, Tianjin, 300071, P.R. China}
	
	\emailAdd{jiakang.bao@city.ac.uk}
	\emailAdd{a.hanany@imperial.ac.uk}
	\emailAdd{hey@maths.ox.ac.uk}
	\emailAdd{edward.hirst@city.ac.uk}
	
	\preprint{
		\begin{flushright}
		    Imperial/TP/21/AH/04
		    
			LIMS-2021-009
		\end{flushright}
	}

	\abstract{Quivers, gauge theories and singular geometries are of great interest in both mathematics and physics. In this note, we collect a few open questions which have arisen in various recent works at the intersection between gauge theories, representation theory, and algebraic geometry. The questions originate from the study of supersymmetric gauge theories in different dimensions with different supersymmetries. Although these constitute merely the tip of a vast iceberg, we hope this guide can give a hint of possible directions in future research. 
		
    This is an invited contribution to a special volume of {\it Proyecciones}, E.~Gasparim, Ed., and it is the hope that the questions are specific enough for research projects aimed at PhD students.
	}

\begin{document}
	\maketitle

\section{Motivation}\label{motivation}
Quivers, introduced to algebraic geometry in \cite{Nakajima:1994nid} and to theoretical physics in  \cite{Douglas:1996sw}, have been a useful tool in the study of supersymmetric gauge theories. In particular, moduli spaces have been extensively studied in the past few decades. The explorations of different branches of the moduli spaces connect various areas in string theory, algebraic geometry, symplectic geometry, tropical geometry, cluster algebra and representation theory.

For instance, brane setups and transitions naturally give rise to these symplectic singularities, especially through affine Grassmannians as shown recently in \cite{Bourget:2021siw}. The notion of magnetic quivers \cite{Cabrera:2019dob} and manipulations of quivers have then been introduced, revealing connections among different theories and geometries. Of late, it has been discussed that superconformal field theories (SCFTs) in higher dimensions, in particular their Higgs branches, can be analyzed via certain spaces of dressed monopole operators. We should also emphasize that the corresponding spaces of dressed monopole operators and magnetic quivers, as a feature of the theories, are studied for their own sake, which in this sense is broader than Coulomb branches of some 3d dual theories.

To analyze the moduli spaces in more details, phase diagrams (aka Hasse diagrams) are introduced to determine the structures of the singularities. On the other hand, Hilbert series (HS) as well as highest weight generating functions (HWGs) have been standard and helpful concepts to enumerate gauge invariant operators (GIOs). Following algebraic geometry, the HS counts GIOs at different degrees, and HWGs provide a more concise notion for HS using highest weight Dynkin labels of the symmetry groups. The HS can be refined or unrefined. When we have multivariables, this gives multi-gradings mathematically and they represent different fugacities for different symmetry groups in our supersymmetric theories. When we perform unrefinement, they would encode less information but have lighter expressions which might be easier to compute.

As a rational function, the HS can be written as a Taylor expansion in the form
\begin{equation}
    \text{HS}=\sum_{k=0}^\infty H(k)t^k,
\end{equation}
where $H(k)$ is known as the Hilbert function. We can also take the Laurent expansion of the (unrefined) HS as
\begin{equation}
	\text{HS}=\sum_{i=0}^\infty\frac{\gamma_i}{(1-t)^{d-i}},
\end{equation}
where $d$ is the (Krull) dimension of the ring. It has been a long-standing problem to understand and determine the Laurent coefficients of the HS. For some special cases, the leading and sub-leading coefficients can be related to the $a$-invariant (i.e., the degree of the numerator minus the degree of the denominator of HS) or even the order and pseudoreflections for a finite group \cite{cowie2019hilbert,herbig2019laurent}. These two coefficients are also shown to be crucial in the study of K-stability \cite{Collins:2012dh,collins2019sasaki}. Physically, the Laurent coefficients can have certain interpretations as well. For instance, $\gamma_0$ gives the normalized volume of the Sasaki-Einstein manifold (e.g. in the context of holography), and hence is related to the study of R-symmetry under RG flows \cite{Martelli:2005tp,Martelli:2006yb}. However, a complete understanding for HS of any ring/variety is still not clear.
\begin{question}
	What are the mathematical and physical interpretations of all the Laurent coefficients of any HS?
\end{question}

\section{Theories with 8 Supercharges}\label{Q8}
Let us begin with rigid theories with 8 supercharges. Such theories have vector multiplets and hypermultiplets. The vector multiplets transform under adjoint reps of the gauge groups while the hypers/matters can transform under any reps. In the context of quivers, we will restrict to (bi)fundamental reps for the hypers\footnote{In 6d with $\mathcal{N}=(1,0)$, there are also tensor multiplets. There are also interesting questions for 6d theories and tensor multiplets, but we will not discuss them here.}. In the language of quivers, we can express the information of the supersymmetric theories with 8 supercharges using
\begin{center}
	\begin{tabular}{|c|m{25em}|}
		\hline
		Round Node $\bigcirc$ & Gauge group and vector multiplet under its adjoint rep                                   \\
		\hline
		Square Node $\square$ & Flavour group                                                                              \\
		\hline
		Line    $\rule[2.5pt]{1cm}{0.05em}$    & Hyper(s) transforming under the bifundamental rep of the groups the line connecting\\
		\hline
	\end{tabular}.
\end{center}
It is also possible to other types of edges in the quivers such as ones with orientations in non-simply laced quivers \cite{Cecotti:2012gh,Cremonesi:2014xha,Chen:2018ntf,Hanany:2020jzl,Bourget:2021xex} and squiggles denoting charge 2 hypers \cite{Bourget:2020gzi}.

\subsection{The Higgs and Coulomb Branches}\label{branches}
In recent years, the 3d $\mathcal{N}=4$ Coulomb branches, realized as spaces of dressed monopoles (to be precise), play a key role in the study of quivers and SCFTs in various dimensions. In particular, various tools have been developed including HS \cite{Cremonesi:2013lqa,Cremonesi:2014kwa} and HWGs \cite{Hanany:2014dia}, Kraft-Procesi (KP) transitions and transverse slices \cite{Hanany:2018uhm,Cabrera:2016vvv,Cabrera:2017njm}, quiver subtractions \cite{Cabrera:2018ann} and quiver additions \cite{Bourget:2021siw}, discrete gauging and quiver origami \cite{Hanany:2018vph,Hanany:2018cgo,Hanany:2018dvd,Bourget:2020bxh}, and magnetic quivers and phase diagrams \cite{Cabrera:2019izd,Bourget:2019aer,Bourget:2019rtl,Cabrera:2019dob,Grimminger:2020dmg,Bourget:2020gzi,Bourget:2020asf,Bourget:2020mez}. There are many interesting perspectives which can be found in these references. Here we will only pick several examples.

The HS is a generating function of a ring/variety/scheme. Physically, it enumerates the GIOs at each degree for the moduli space (which will mainly be the Higgs or Coulomb branch here in our discussion). Importantly, magnetic quivers (as fundamental properties of the higher dimensional theories) allow us to formally equate Coulomb branches for certain 3d theories with Higgs branches for higher dimensional theories as moduli spaces. Therefore, we are able to analyze the moduli spaces in 5d or 6d using the knowledge of 3d theories\footnote{As aforementioned, we should really treat this as a property of the 5d or 6d theory itself.}. Since they are \emph{symplectic singularities}, they are composed of \emph{symplectic leaves} and these leaves form a poset by inclusions of their closures. Every smaller leaf (in the sense of the partial order) has a \emph{transverse slice} to it in a larger leaf. Every transverse slice can be viewed as a moduli space for some theory which can be obtained from (partial) Higgsing from the unbroken theory.

On the other hand, it is convenient to use phase diagrams to encode the poset structure of the symplectic singularities. In a phase diagram, the nodes are often labelled by the (quaternionic) dimensions of the leaves and the edges have numbers denoting the dimensions of the neighbouring transverse slices (aka \emph{elementary slices}). Going from the bottom of the phase diagram to the top is the process of Higgsing the theory, where the top corresponds to the fully Higgsed theory at the origin of the full Higgs branch while each segment starting from the top to some node in the middle (or at the bottom) corresponds to certain effective theory where the gauge group is partially (or completely) broken.

With these two powerful tools, it would be natural to ask the following question.
\begin{question}
	Can we compute the phase diagrams from the HS/HWGs?
\end{question}
The phase diagrams for the full moduli spaces (including mixed branches) were studied in \cite{Grimminger:2020dmg}, and it was found that certain theories have Higgs and Coulomb branches which are related by \emph{inversion} of their phase diagrams. On the other hand, a duality known as \emph{symplectic duality} between Higgs and Coulomb branches of a 3d $\mathcal{N}=4$ theory was conjectured in \cite{braden2014quantizations} and has been extensively studied in \cite{Nakajima:2015txa,nakajima2015questions,Braverman:2016wma,webster2016koszul,Bullimore:2016nji,Balasubramanian:2018pbp,Dancer:2020wll,Bourget:2021zyc}. Therefore, for phase diagrams, another question could be raised.
\begin{question}
	Can we find any relation between symplectic duality and inversions of phase diagrams?
\end{question}
All the related questions either mentioned or not mentioned here would help us get a better understanding of the moduli space. Besides, there is a fundamental question we should always mention:
\begin{question}
	What is the mathematical definition of Coulomb branches (for 3d $\mathcal{N}=4$ theories)?
\end{question}
Nakajima and collaborators have quite a few inspiring works on this including \cite{Nakajima:2015txa,nakajima2015questions,Braverman:2016wma,Nakajima:2017bdt}. This problem would be of great interest to both mathematicians and physicists.

Recently, Nakajima and Weekes have also generalized this to more general symmetrizable theories \cite{Nakajima:2019olw}. In particular, quiver folding allows one to get non-simply laced quivers from simply-laced ones. This relates Coulomb branches for unfolded and folded quivers via finite group actions. For these non-simply laced quivers, the HS and HWGs have been calculated in \cite{Bourget:2020bxh} for unitary quivers and in \cite{Bourget:2021xex} for orthosymplectic quivers. However, it is still not clear on how to compute the Higgs branches.
\begin{question}
    Can we compute HS and HWGs for the Higgs branches of folded theories?
\end{question}

\subsection{Canonical Singularities and 5d SCFTs}\label{canonical}
Besides the techniques introduced above, a parallel study for 5d $\mathcal{N}=1$ SCFTs appeared in \cite{Xie:2017pfl}. In that work, Xie and Yau conjectured that every three dimesional canonical singularities would give rise to a 5d SCFT with $\mathcal{N}=1$. In particular, the chamber structure and prepotential of the Coulomb branch are studied using the Nef cones from crepant resolutions.

Following the works \cite{Closset:2020scj,vanBeest:2020kou,vanBeest:2020civ}, \emph{dot diagrams} (aka \emph{generalized toric polygons}) which were first introduced in \cite{Benini:2009gi} were used to study those 5d $\mathcal{N}=1$ SCFTs. The aforementioned tools including magnetic quivers and phase diagrams can also be successfully applied under various manipulations of the dot diagrams. Therefore, one may raise the following question.
\begin{question}
	Does M-theory on any 3d canonical singularity always define a 5d $\mathcal{N}=1$ SCFT (conjecture 1 in \cite{Xie:2017pfl})? Can we use the tools including dot diagrams and magnetic quivers to verify this conjecture?
\end{question}

Incidentally, the \emph{(generalized) s-rule} can also be described in terms of dot diagrams following \cite{Benini:2009gi,vanBeest:2020kou}. It is straightforward to incorporate the s-rule using tessellations of the polygon, but it is hard to write a more sufficient algorithm in the sense of coding (and hence there is a strong s-rule in \cite{vanBeest:2020kou}) so that none of the possible tessellations would be missed. On the other hand, the s-rule can be determined by the self-intersection number of the holomorphic curve that M2 wraps for irreducible brane junctions \cite{DeWolfe:1998eu,Bergman:2020myx}.
\begin{question}
	Can we write down the criterion of generalized s-rule in terms of self-intersection numbers for reducible junctions?
\end{question}
\begin{question}
	Is there a faster way/(coding) algorithm to determine generalized s-rule using dot diagrams?
\end{question}

\subsection{Affine Grassmannians}\label{affgr}
In a recent paper \cite{Bourget:2021siw}, the \emph{affine Grassmannian} (which is an ind-scheme) has been introduced to the study of brane setups in string theory, where Coulomb branches (for framed quivers) are slices in it \cite{Braverman:2016pwk}. The authors worked out the transverse slices and phase diagrams for different affine Grassmannians by studying the corresponding brane systems and quivers. Here, we will only give the definition of affine Grassmannians. See \cite{Bourget:2021siw} for detailed calculations and examples.
\begin{definition}
	Let $k$ be a separably closed field. The ring of formal power series and the formal Laurent series over it are denoted as $k[[t]]$ and $k((t))$ respectively. The affine Grassmannian $\textup{Gr}_G$ of a connected reductive group $G$ is the coset space $G(k((t)))/G(k[[t]])$.
\end{definition}
In \cite{Bourget:2021siw}, the authors focused on $k=\mathbb{C}$ and any finite dimensional Lie group $G$. In this case, $\text{Gr}_{G}$ is the quotient $G(\mathbb{C}((t)))/G(\mathbb{C}[[t]])$. Equivalently, it is defined to be the set of all \emph{lattices} in $k((t))^n$, where a lattice in $k((t))^n$ is a free $k[[t]]$-module of rank $n$.

In the language of category theory, the definition can be reformulated as follows \cite{richarz2016affine}. Let $\bm{k}$\textbf{-Algebra} and \textbf{Grp} be the category of commutative $k$-algebra and the category of (small) groups. The \emph{loop group} is the group functor $L_tG:\bm{k}\textbf{-Algebra}\rightarrow\textbf{Grp},~A\mapsto G(k(t))$. Likewise, the \emph{positive loop group} is the group functor $L_t^+G:\bm{k}\textbf{-Algebra}\rightarrow\textbf{Grp},~A\mapsto G(k[[t]])$. Then the affine Grassmannian is the (fpqc-)quotient $\text{Gr}_G=L_tG/L_t^+G$.

\begin{question}
	Study the Coulomb branch for ``ugly'' and ``bad'' theories (in the sense of \cite{Gaiotto:2008ak}) as (generalized) affine Grassmannian slices.
\end{question}
As discussed above, besides the Coulomb branches of the quivers, the Higgs branches can be obtained from inversions of phase diagrams.
\begin{question}
	Can we have some notion of symplectic duals for the affine Grassmannians?
\end{question}

The unitary quivers without loops for elementary slices including those in the study of minimal nilpotent orbits \cite{Hanany:2016gbz,Hanany:2017ooe,Eager:2018dsx} and in the study of affine Grassmannian are listed in \cite[Table 1]{Bourget:2021siw}. Nevertheless, the list is still incomplete. Therefore, a classification of elementary slices (including those not from affine Grassmannians) is still not known.
\begin{question}
    Classify all the elementary slices.
\end{question}

It is also worth noting that the notion of quiver addition was also introduced in \cite{Bourget:2021siw}. Together with quiver subtractions, we may ask:
\begin{question}
	Can we give a rigorous mathematical description of quiver subtractions and quiver additions?
\end{question}

\section{Theories with 4 Supercharges}\label{Q4}
Next, we move on to theories with fewer supersymmetries, viz, with four supercharges which exist in dimension less than or equal to 4. In particular, we will mainly focus on 4d $\mathcal{N}=1$ theories here whose quivers can be obtained from decomposing $\mathcal{N}=2$ multiplets using the following rules:
\begin{center}
	\begin{tabular}{|c|c|}
		\hline
		4d $\mathcal{N}=2$ & 4d $\mathcal{N}=1$  \\ (8 supercharges) & (4 supercharges)                               \\
		\hline
		Round Node  & Round Node with Directed Loop  \\
		\begin{tikzpicture}
			[round/.style={circle, draw=black, fill=white, minimum size=2mm}
			]
			\node[round] (a) at (0,0){};
			\node (b) at (0,-0.25){};
		\end{tikzpicture} &
		\tikzset{every picture/.style={line width=0.75pt}}
		\begin{tikzpicture}[x=0.75pt,y=0.75pt,yscale=-1,xscale=1]
\draw   (100,127.65) .. controls (100,123.43) and (103.43,120) .. (107.65,120) .. controls (111.87,120) and (115.3,123.43) .. (115.3,127.65) .. controls (115.3,131.87) and (111.87,135.3) .. (107.65,135.3) .. controls (103.43,135.3) and (100,131.87) .. (100,127.65) -- cycle ;
\draw    (114.3,134.2) .. controls (141.32,141.92) and (138.54,113.32) .. (115.85,120.3) ;
\draw [shift={(113.3,121.2)}, rotate = 338.2] [fill={rgb, 255:red, 0; green, 0; blue, 0 }  ][line width=0.08]  [draw opacity=0] (10.72,-5.15) -- (0,0) -- (10.72,5.15) -- (7.12,0) -- cycle    ;
\end{tikzpicture}
		\\ Vector multiplet  &    Vector multiplet \& Adjoint Chiral Multiplet                                                                        \\
		\hline
		Line      &  Bidirectional Line\\
		$\rule[2.5pt]{1cm}{0.05em}$ &
		\begin{tikzpicture}
			[round/.style={circle, draw=black, fill=white, minimum size=2mm}
			]
			\tikzstyle{arrow} = [,<->,>=stealth]
			\node (a) at (0,0){};
			\node (b) at (1,0){};
			\draw[arrow] (a)--(b);
		\end{tikzpicture}
		\\ Hypers  &      Chiral and Anti-Chiral Multiplets          \\
		\hline
	\end{tabular}.
\end{center}
For 3d $\mathcal{N}=2$ theories, the quivers can be obtained from $\mathcal{N}=4$ theories with the same rules. As we can see, the quivers for 4d and 3d theories with 4 supercharges share some common features. Both of them have R-symmmetry U(1), yet the underlying physics could still be quite different.

\subsection{Minimized Volumes and Their Bounds}\label{minvol}
A very-well studied class of 4d theories is the worldvolume theories of D3-branes probing toric Gorenstein singularities \cite{Feng:2000mi,He:2001ey,Feng:2004uq,Feng:2001bn,Hanany:2005ve,Franco:2005rj,Bianchi:2021uhn,Eager:2010ji,craw2008quiver,broomhead2009dimer,bocklandt2016dimer,Cachazo:2001gh,Cachazo:2001sg,Yamazaki:2008bt}\footnote{In literature, this is often referred to as Calabi-Yau (CY) $n$-folds. However, to be strict, we will save the name CY for compact (smooth) manifolds and call the unresolved singularities Gorenstein. For other types of branes probing Gorenstein singularities, see for example \cite{Gukov:2002es,Fluder:2015eoa,Hanany:2008cd,Hanany:2008fj,Hanany:2009vx,Davey:2009bp}.}. The forward and inverse algorithms allow us to go between toric diagrams and quivers as well as brane tilings easily. The holographic dual of such a theory would live in AdS$_5\times Y_5$ where the Sasaki-Einstein (SE) manifold $Y_5$ (of real dimension 5) is exactly the base of the 3d toric Gorenstein singularity. Even more remarkably, this SE base has a close relation to the determination of R-symmetry of our 4d theory. Under RG trajectory, the R-symmetry at IR fixed point may become a linear combination of the original U(1)$_R$ and some abelian flavour symmmetries preserved through the flow. To determine the R-charges of the operators in our theory, Intriligator and Wecht proposed a procedure known as the $a$-maximization in \cite{Intriligator:2003jj} where $a$ is one of the central charges in 4d. The idea that $a$ decreases along RG flow was first conjectured by Cardy in \cite{Cardy:1988cwa} and was proven over two decades later by Komargodski and Schwimmer in \cite{Komargodski:2011vj}.

For the worldvolume theories of D3s discussed here, Gubser showed in \cite{Gubser:1998vd} that $a=\frac{1}{4V_n}$ where $V_n$ is the volume function of the SE base $Y$ normalized by the volume of the 5-sphere\footnote{In general for $n$-dimensional Gorenstein, $V_n$ is the normalized volume for $Y_{2n-1}$, viz, $\text{vol}(Y_{2n-1})/\text{vol}(S^{2n-1})$.}. Therefore, $a$-maximization has now been translated to the problem of \emph{volume minimization} for Sasaki-Einstein manifolds. See also \cite{Eager:2010yu,Butti:2005vn}.

From a lattice polygon, we not only have a non-compact 3d toric Gorenstein singularity, but can also construct a compact toric variety (which is not necessarily CY) from the inner normal fan of the polygon. In general, the discussions so far do not need to be restricted to 3d Gorenstein singularities but any dimension $n$. Every $Y_{2n-1}$ still has a minimized volume and we can always construct some compact toric variety $X$ of complex dimension $(n-1)$ from the corresponding $(n-1)$-tope. Physically, they may also have some string theory picture in terms of different D$p$-branes in certain dimensions.

In \cite{He:2017gam,Bao:2020kji}, systematic computations of minimized volumes were performed. In particular, one can relate the minimized volumes to the topological invariants of $\tilde{X}$ where the tilde denotes the \emph{complete} (crepant) resolution of $X$, as $X$ may not be smooth\footnote{Following \cite{batyrev1982toroidal,nill2005gorenstein}, $X$ whose dimension is no less than 4 may not be completely desingularized. In such cases, we will only focus on toric varieties that admit complete resolutions.}. In particular, we shall always take \emph{fine} triangulations, that is, triangulations involving all the lattice points. The authors conjectured that the minimized volumes have certain bounds in terms of the Chern numbers of $X$. This then naturally relates geometric and topological quantities for different objects. The most updated conjecture can now be written as follows.
\begin{conjecture}
	For an $n$-dimensional toric Gorenstein singularity associated with a polytope $\Delta_{n-1}$ (either reflexive or non-reflexive), the minimized volume $V_{n,\min}$ of the SE base manifold $Y$ is bounded by
	\begin{equation}
		\frac{1}{\chi}\leq V_{n,\min}<m_n\int c_1^{n-1},\label{bd1}
	\end{equation}
where $\chi$ and $c_1$ are the Euler number and first Chern class of the complete resolution $\tilde{X}$. Moreover, the left bound is saturated when the Gorenstein singularity is an Abelian orbifold of $\mathbb{C}^n$.

In particular for \emph{reflexive} cases, the coefficients $m_n$ are conjectured to be positive and satisfy $m_3\sim3^{-3}$, $m_4\sim4^{-4}$ and $m_n>m_{n+1}$.
\end{conjecture}\label{Vminconj}
It is worth noting that such bounds for 2d lattice polygons in terms of their areas are obtained in \cite{Gulotta:2008ef}:
\begin{equation}
    \frac{1}{A}\leq V_{3,\min}<\frac{4\pi^2}{27A},\label{bd2}
\end{equation}
where $A$ is the normalized area of the polygon. Moreover, the lower bound is saturated for triangles while the upper bound is the case for ellipses (as limit shapes of polygons) and hence can never be saturated. If we compare \eqref{bd2} with \eqref{bd1} with $n=3$, we find that the lower bounds agree: $\chi$ is the Euler number for a complete resolution which corresponds to a fine triangulation, and hence $\chi=A$. Furthermore, they take equalities under the same condition. For the upper bound, \eqref{bd1} becomes $m_3C_1=m_3(12-\chi)$, which is more subtle to understand its connection to \eqref{bd2}. This leads to the question below.
\begin{question}
	Prove or disprove Conjecture \ref{Vminconj}. If the conjecture is true, are these the best bounds? If the conjecture is not true, find the correct bounds. Also, can we find the connections between \eqref{bd1} and \eqref{bd2} (in particular for their upper bounds)?
\end{question}
Furthermore, conjecture 5.4 in \cite{He:2017gam} gives rise to another question for minimized volumes and Euler numbers. In particular, the authors observed that maximum value of $V_{n,\min}\chi$ is reached at dP$_3$ for $n=3$ and some fibrations of dP$_3$ for $n=4$ \cite[Figure 15]{He:2017gam}. One may then ask:
\begin{question}
	Is the maximum value for $V_{n,\min}\chi$ for reflexive polytopes attained by various (not necessarily uniquely) \textup{dP}$_3$ fibrations?
\end{question}

\subsection{K-Stability and Chiral Rings}\label{kstability}
As discussed above, $a$-maximization is closely related to volume minimization. Following \cite{Collins:2012dh,collins2019sasaki}, \emph{K-stability} naturally appears in the study of $a$-maximization. More specifically, the authors showed that K-(semi)stability for product test configurations is equivalent to volume minimization. Therefore, in \cite{Collins:2016icw}, K-stability for general test configurations was regarded as some ``generalized $a$-maximization''.

Given a chiral ring $I$ which is a ring composed of chiral operators under operator product expansions, we can compute the HS of the associated variety $X=\text{Spec}(I)$. Given the (unrefined) HS of some $X$ graded by the one-dimensional symmetry\footnote{Physically, $\zeta$ is treated as our U(1) R-symmetry.} $\zeta$, the strategy is to ``perturb'' it with some test symmetry $\eta$ such that the new HS is graded by $\zeta+\epsilon\eta$ for sufficiently small $\epsilon>0$. Then one can compute the \emph{(Donaldson-)Futaki invariant} for this test symmetry using the (leading and subleading) coefficients in the Laurent expansions of HS. Now $X$ is said to be K-semistable if $F\geq0$ for any test symmetry $\eta$. For it to be K-stable, $F$ can be zero only when the norm vanishes. For the expression of Futaki invariants and the definition of norm, one is referred to \cite{szekelyhidi2014introduction,Bao:2020ugf}. More details of defining K-stability along with its calculations can also be found in \cite{Bao:2020ugf}.

In \cite{Collins:2016icw}, it was then conjectured that a ring is the chiral ring for an SCFT iff $X$ is K-stable. Indeed, for some theories (such as the worldvolume theory of D3s probing Gorenstein singularties), one can show that K-stability recovers the unitarity bounds or irrelevance of superpotential terms \cite{Collins:2016icw}. However, a counterexample was found in \cite{Bao:2020ugf}. Therefore, it is natural to ask:
\begin{question}
	Can we give a precise description on how K-stability is related to chiral rings and SCFTs?
\end{question}

Computationally, it is still not known how to fully determine K-stability for a general variety as one needs to check infinitely many test symmetries in principle. For hypersurface singularities, especially for those having U(1)$^{n-1}$ isometry, not only the number but also exactly which test symmetries can be determined to compute Futaki invariants. See \cite{ilten2017k,Fazzi:2019gvt} for more details. Nevertheless, the problem for a general variety is still unsolved.
\begin{question}
	How to determine the number of test symmetries one needs to check to determine K-stability for a general variety? Can we further determine exactly which test symmetries should be checked?
\end{question}

Of course, it is reasonable to wonder whether there is any analogue in other dimensions, such as $F$-theorem in 3d or $c$-theorem in 2d etc.
\begin{question}
	Can we also apply K-stability to chiral rings in other dimensions? Can K-stability cope with accidental symmetries that might appear under RG flow?
\end{question}
Moreover, Benvenuti and Giacomelli introduced another chiral ring stability in \cite{Benvenuti:2017lle} in terms of dropping certain terms in the superpotential.
\begin{question}
	Is there any relation between the two stabilities for chiral rings?
\end{question}

\subsection{Graded Quivers: from $m=1$ to general $m$}\label{graded}
Now let us relax the restriction of fixed number of supercharges and take a quick look at \emph{graded quivers}, aimed at providing a unified mathematical framework for gauge theories in even dimensions. Graded quivers have been extensively studied in \cite{Franco:2017lpa,Closset:2018axq,Franco:2019bmx,Franco:2020ijt,Franco:2020avj}. As the name suggests, the arrows/fields in an $m$-graded quiver are associated with a grading by some quiver degree $c\in\{0,1,\dots,m\}$. These different types of fields are denoted by $\Phi^{(c)}_{ij}$ for an arrow pointing from node $i$ to node $j$ with $c$ ranging from 0 to $\lfloor m/2\rfloor$. When a field has degree $m/2$ (for even $m$), it is unoriented. For those of degree $(m-c)$, we can use the notion of conjugate arrow with an opposite direction: $\bar{\Phi}^{(m-c)}_{ji}\equiv\overline{\Phi^{(c)}_{ij}}$. Physically, such conjugation refers to CPT conjugate fields. Superpotentials, $W$, are always crucial for a graded quiver. They are linear combinations of gauge invariant terms of degree $(m-1)$ satisfying $\{W,W\}=0$, where $\{\text{-},\text{-}\}$ is the \emph{Kontsevich bracket} which is a generalization of the Poisson bracket \cite{Franco:2017lpa}.

In the string theory picture, graded quivers generalize the D3s probing 3d Gorenstein singularities to D($5-2m$)-branes transverse to $(m+2)$-dimensional Gorenstein singularities in Type IIB string theory. Usually, we consider $m=0,1,2,3$ whose theory has $2^{3-m}$ supercharges in dimension $(6-2m)$ (see for example the table in (1.2) in \cite{Closset:2018axq}).

Analogous to Seiberg duality in 4d, graded quivers also enjoy certain dualities in other dimensions such as triality in 2d and quadrality in 0d \cite{Franco:2020ijt}. Moreover, the generalized brane tilings on $\mathbb{T}^{m+1}$ for $m$-graded quivers have also been developed in \cite{Franco:2019bmx}. More recently, a notion of product of toric diagrams was introduced in \cite{Franco:2020avj} which allows one to produce the quiver theory for an $(m+n+3)$-dimensional Gorenstein singularity from a pair of theories for $(m+2)$- and $(n+2)$-dimensional Gorenstein singularities. Here we list some of the questions in the study of graded quivers.
\begin{question}
	Study the dualities for graded quivers involving adjoint fields.
\end{question}
\begin{question}
	For brane tilings on $\mathbb{T}^2$, zig-zag paths and perfect matchings are important concepts. Can we reveal their further mathematical and physical implications for general $m$?
\end{question}
\begin{question}
	Graded quivers can be obtained from the topological B-model. Can we find any correspondence between the product of toric diagrams and the B-model approaches?
\end{question}

\subsection{Dessins d'Enfants}
Dessins d'enfants are bipartite graphs, whose primary role is in the study of Riemann surfaces $X$. Specifically they represent the degeneracies and ramifications of so-called Bely\v{\i} maps which take $X$ to the Riemann sphere $\mathbb{P}^1$ \cite{Grothendieck}.
They received particular popularity due to the faithful action on them by the absolute Galois group over the rational numbers: Gal$(\overline{\mathbb{Q}}/\mathbb{Q})$, whereby making them a point d'appui in geometry and number theory.

Let us begin with the classic theorem of Bely\v{\i} \cite{Bely__1980}.
\begin{theorem}[Bely\v{\i}]
A Riemann surface $X$ has an algebraic model over $\overline{\mathbb{Q}}$  IFF there exists a (surjective) map $\beta : X \rightarrow \mathbb{P}^1$ which is ramified at exactly 3 points. 
\end{theorem}
In the theorem, $\beta$ is the Bely\v{\i} map, and after M\"obius transformation on the $\mathbb{P}^1$, the three ramification points can be mapped to: $\{0,1,\infty\}$.
The surface together with the map, $(X,\beta)$, is called the Bely\v{\i} pair.

The components of the dessin d'enfant are then identified with these points via:
\begin{equation}
    \beta^{-1}(0) \longmapsto \circ\;,\quad \beta^{-1}(1) \longmapsto \bullet\;,\quad \beta^{-1}((0,1)) \longmapsto -
\end{equation}
such that the preimages of the $[0,1]$ interval on the Riemann surface, $X$, form the dessin d'enfant, and in addition the preimages of $\infty$ are associated to the dessin d'enfant's faces.

\subsubsection{Brane Tilings and Modular Parameters}
The bipartite nature of dessin d'enfants allows for a natural association to brane tilings (\textit{a.k.a.} dimer models). Brane tilings encode quiver gauge theories with toric moduli spaces as bipartite graphs drawn on the torus, $\mathbb{T}^2$ \cite{Hanany:2005ve,Franco:2005rj,Feng:2005gw,Yamazaki:2008bt}; the bipartite nature of these theories are considered in
\cite{Jejjala:2010vb,Heckman:2012jh,Franco:2012mm,Bianchi:2014qma} and the relation to amoebae and tropical geometry, where the dual of the toric diagrams are so-called brane webs \cite{Aharony:1997bh} are considered in \cite{Feng:2005gw}.

Brane tilings are an alternative graphical method of representing the information of a quiver gauge theory with superpotential. 
They are related to the quiver dual graph, whereby tiling faces associate to U$(N)$ gauge groups, tiling edges to chiral multiplets, and tiling vertices to superpotential terms. The bipartite nature of the tiling identifies the orientation about which to write the vertex's incident edges for the vertex's monomial term in the superpotential. 
Interpreting genus 1 dessins d'enfants as brane tiling bipartite graphs, and vice versa, lead to a series of intriguing questions about the connections between them.

Quite interestingly a specific limit of the brane tiling's field theory can be graphically represented in a specific drawing of the brane tiling \cite{Hanany:2011ra}. In the SCFT, performing $a$-maximisation for the central charge $a$ as a function of the fields' R-charges, fixes the latter.
These R-charge values are encoded in the tiling through {\it isoradial embedding}; which draws all nodes as lying on a unit circle centred on the face they border, whilst at the centre the angle subtended by a bordering edge corresponding to R-charge $R_i$, is $\pi (1-R_i)$.
The field theory maximisation is performed under the following conditions: (i) all superpotential terms maintain an R-charge of 2 (s.t.~Lagrangian in superspace is well-defined) corresponding geometrically to the face centre's total angle equaling $2\pi$; and (ii) $\beta$-functions vanishing (to ensure conformality) corresponding to the faces forming a rhombus tiling \cite{Hanany_2007}.
There is a notion of ``consistency'' of whether these conditions suffice physically and mathematically \cite{Davey:2009bp,broomhead2009dimer,Franco:2017jeo}, and a question emerges as to
\begin{question}
Given an inconsistent tiling, what is the infra-red physics, and the geometry of the moduli space?
On the other hand, for consistent theories, it would be important to check that the renormalization group flow in the field theory takes one to the correct geometry.
\end{question}

An isoradial brane tiling dictates a specific form of the underlying torus.
Its modular parameter, and hence, complex structure, can be extracted. We denote it as $\tau_R$ (to emphasize the R-charge origin).
On the other hand, the Bely\v{\i} pair fixes the complex structure of $X$, which we denoted as $\tau_B$ (to emphasize the Bely\v{\i} origin).

The equality of these two complex structures was first conjectured in \cite{Jejjala:2010vb}, and further work has found an array of examples where this equality holds ($\mathbb{C}^3$, conifolds, and their orbifolds).
However, a counter-example in
\cite{Hanany:2011ra}, viz., $L^{2,2,2}$, showed that these two complex structures need not be equal in general.

Now, Seiberg duality preserves $\tau_R$ from a physical point of view, as shown in \cite{Hanany:2011bs}.
However, its action on $\tau_B$, and indeed on the dessin, is not clear.
These discussions naturally lead to questions about the scope of the equality and the parameters' respective interpretations.
\begin{question}
    In what scenarios does $\tau_B = \tau_R$ hold?
\end{question}
\begin{question}
    What is the physical interpretation of $\tau_B$ in the brane tiling's SCFT?
\end{question}
\begin{question}
    Does $\tau_R$ have use in the theory of dessins d'enfants, or further in Galois theory?
\end{question}

Another modular parameter also appears in this context, this time associated to a torus arising directly from the geometry.
In the cases where the quiver gauge theory's toric vacuum moduli space is Gorenstein in nature, this Gorenstein singularity can be reformulated as a torus fibration.
From there, the action of the 3d mirror symmetry acts alike a T-duality with U$(1)^3$ symmetry \cite{Strominger_1996, Feng:2005gw}, however only a U$(1)^2$ subgroup preserves the K\"ahler form and holomorphic 3-form. 
This U$(1)^2$ may be used to define an invariant part of the Gorenstein's SLag (special Lagrangian), such that this part will be a torus.
Finally, pulling-back the metric to this torus where the brane tiling exists provides a metric.
This torus' metric's complex structure is then another modular parameter, denoted $\tau_G$ to emphasise the geometric motivation \cite{He:2012xw}.

It is then instinctive to ask where this $\tau_G$ fits in with the other modular parameters.
Computations in \cite{He:2012xw} showed that $\tau_G = \tau_R$ in some simple cases ($\mathbb{C}^3$, conifold); whilst also that this equality holds only approximately in more complicated scenarios ($L^{1,2,1}$). 

\begin{question}
    What is the precise relation between $\tau_G$, $\tau_R$ and $\tau_B$?
\end{question}

\subsubsection{Galois Orbits and Seiberg Duality}
The dessin d'enfant Bely\v{\i} maps, through Bely\v{\i}'s theorem, may be defined using algebraic numbers exclusively.
However which specific field extension of the rationals is sufficient to describe them depends on the Bely\v{\i} pair in question. 
The roots used in the Galois extension can be considered as roots of some minimal separable polynomial over the rationals.
Changing which of these roots is used in the field extension for the Bely\v{\i} map takes the Bely\v{\i} map, and respective dessin, through its Galois orbit.
Through Galois orbits dessins manifestly group themselves, and how these orbits find use physically is an area of particular interest.

Contrastingly, on the physical side quiver gauge theories naturally sort themselves into duality classes, based on the action of Seiberg duality \cite{Seiberg_1995,Feng_2001,Franco_2004,Franco:2003ja}.
Seiberg duality is an IR equivalence between 4d $\mathcal{N}=1$ theories, in our case this importantly also applies to toric quiver gauge theories.
The brane tilings for these theories thus form groupings as bipartite graphs within the same duality trees, alike the groupings of dessins d'enfants under Galois conjugation.
How these two groupings arise in each others fields is suggested to have some interesting significance.
\begin{question}
    Is there physical significance to the relation between brane tiling theories whose bipartite graphs as dessins d'enfants are in the same Galois orbit?
\end{question}
\begin{question}
    How does Seiberg duality between brane tilings relate dessins d'enfants, and what Galois invariants (if any) can be inspired by these duality trees?
\end{question}
Since a primary goal within Galois theory is identifying Galois invariants that can be used as tools in calculation, perhaps Seiberg duality trees may reveal some physically-inspired invariants with use on the mathematical side.

\subsubsection{Seiberg-Witten Curves and Modular Surfaces}
Seiberg-Witten curves are exact descriptions of the Coulomb branch for the IR limit of $\mathcal{N}=2$ gauge theories.
Deforming these theories with tree-level superpotentials causes symmetry breaking to $\mathcal{N}=1$ theories, where vacua produced are connected in phases based on the deforming superpotential's parameters \cite{Cachazo_2003}.

Through tuning the deforming superpotential's parameters, the roots of the components of the $\mathcal{N}=2$ hyperelliptic curve can be shown to coalesce, such that a drawing of the roots connected by the relevant branch cuts forms a bipartite graph. 
These bipartite graphs occur in the moduli space exclusively at the points the Seiberg-Witten curve develops isolated singularities, considered special points of the $\mathcal{N}=1$ phases.

These bipartite graphs are dessins d'enfants, and it was shown in \cite{Ashok:2006br} that order parameters separating different branches of the $\mathcal{N}=1$ vacua may be considered as Galois invariants. 
This surprisingly link between the vacua phases of this symmetry breaking and dessins d'enfants inspires curiosity into the deeper connection between these objects.
\begin{question}
    How are $\mathcal{N}=1$ vacua from Seiberg-Witten curves and dessins d'enfants explicitly linked?
\end{question}
Their association would lead to powerful insight on both sides, identifying further Galois invariants, as well as deeper physical meaning. 
In addition to the relation to Seiberg-Witten hyperelliptic curves, dessins d'enfants also arise in relation to elliptic curves in another way.

Modular curves are formed through quotient action on the upper half plane, $\mathcal{H}$, by subgroups of the modular group, PSL$(2,\mathbb{Z})$. After extension of this modular action to $\mathcal{H} \times \mathbb{C}$, the `modular surfaces' formed for the index 24 subgroups are surprisingly K3 surfaces, and those for index 36 subgroups are Calabi-Yau 3-folds \cite{He_2013}.

Examining the elliptic $j$-invariants for these surfaces in elliptic Weierstra{\ss} form, the fibration over $\mathbb{C}$ from the modular action extension turns these invariants into maps. 
These maps astoundingly turn out to be Bely\v{\i} in nature, and thus these elliptic surfaces are related to dessins d'enfants quite naturally. 
Moreover, the same dessins can be seen to arise from these subgroups more directly through their Schreier coset graphs \cite{He_2013a}.

This unexpected connection between modular surfaces and dessins d'enfants through the modular group stimulates further questions about its scope for other modular subgroups.
\begin{question}
    To what extent do modular subgroups connect modular surfaces to dessins d'enfants?
\end{question}

\section{A Digression: Machine Learning}\label{ML}
In recent years string and gauge theories have capitalised on a modern computational tool: machine-learning (ML). 
Its use initiated in this field with the examination of string landscapes \cite{He:2017aed,Krefl:2017yox,Ruehle:2017mzq,Carifio:2017bov}, and has since quickly developed these techniques to a wide range of subfields related to gauge theories.
In particular, ML has seen great success in topics discussed in this paper, examining: plethystics \cite{Bao:2021auj,Bao:2021aaa}, amoebae \cite{Bao:2021olg}, Seiberg duality \cite{Bao:2020nbi}, and dessins d'enfants \cite{He:2020eva}.

ML finds its use within this field in two primary scenarios. 
The first being for speed in computation, since many problems have beyond polynomial time complexity to solve. Where these problems require minimal computation to check a solution, ML tools can quickly, and computationally cheaply, provide an array of predicted solutions which can easily be confirmed if valid or not.
The second scenario is in conjecture formulation, the complexity of many ML tools makes them excellent at higher-dimensional pattern recognition, particularly useful when examining large datasets to spot relations and aid in forming conjectures.

The field of study of ML separates itself into 3 styles of problem: supervised, unsupervised, and reinforcement.
Supervised learning uses tools such as neural networks, support vector machines, random forests to perform non-linear function fitting between known inputs and outputs. These tools are used with the aim that the function will have predictive power for both interpolation and extrapolation beyond the training dataset.
Unsupervised learning uses tools such as clustering and autoencoders to identify patterns in data, and isolate degrees of freedom.
Finally reinforcement learning trains an agent to efficiently search for optimum/desired solutions within a known state space, in a Markov Decision Process style.

Neural networks often use the ReLU function as the non-linear component, s.t. ReLU$(x) \vcentcolon= \max(x,0)$.
The linear combination of these functions between neural network neurons makes the full function used to approximate the supervised data piecewise linear in nature.
Interestingly, and surprisingly developed independently, the combination of linear action and maximisation finds itself at the heart of another field also, tropical geometry.
Only recently has work been initiated to examine the interrelation between these areas \cite{zhang2018tropical}, and calls for further exploration.
\begin{question}
	Can we establish a complete correspondence between tropical geometry and neural networks?
\end{question}

Since there is a natural connection between tropical geometries through amoebae and quiver gauge theories, the previous question would link into our primary topic:
\begin{question}
    Can the tropical functions describing ReLU NNs be redescribed as quiver gauge theories?
\end{question}

The action of neural network training has recently drawn inspiration from traditional quantum field theory techniques. 
The non-gaussianity of the training process has been related to renormalisation group flow, with different trained networks acting alike fixed points in the flow \cite{Koch:2019fxy,Halverson:2020trp} (q.v.~\cite{Hashimoto:2019bih}).
Cementing these ideas is an interesting topic for further work.
\begin{question}
    Can the success of machine learning methods be explained through quantum field theory techniques?
\end{question}

\section*{Acknowledgement}
We would like to thank Antoine Bourget, Julius Grimminger, Ali Zahabi, Zhenghao Zhong for enjoyable disscussions and clear explanations on their works. JB is supported by a CSC scholarship. The work of AH is supported by STFC grant ST/P000762/1 and ST/T000791/1. YHH would like to thank STFC for grant ST/J00037X/1. EH would like to thank STFC for the PhD studentship.

\linespread{0.5}\selectfont

\addcontentsline{toc}{section}{References}
\bibliographystyle{utphys}
\bibliography{references}

\end{document}